\documentclass[conference]{IEEEtran}
\usepackage{epsfig}
\usepackage{graphicx}
\usepackage{subfigure}

\newtheorem{definition}{Definition}
\newtheorem{corollary}{Corollary}

\usepackage{amssymb}
\usepackage{makeidx}
\usepackage{amsmath}
\usepackage{graphicx}
\usepackage{epsf}
\usepackage{epsfig}
\usepackage{psfig}
\usepackage{ccaption}
\usepackage{array}
\usepackage{tabularx}
\usepackage{multirow}
\usepackage{epsfig}
\usepackage{cite}
\usepackage{color}

\long\def\symbolfootnote[#1]#2{\begingroup%
\def\thefootnote{\fnsymbol{footnote}}\footnote[#1]{#2}\endgroup}

\begin{document}
\topmargin=-0.6in \oddsidemargin -0.45in \textwidth=7.2in
\textheight=9.7in

\title{\huge{Crystallized Rates Region of the Interference Channel via Correlated Equilibrium with Interference as Noise}}
\author{Mohamad Charafeddine, Zhu Han$^*$, Arogyaswami Paulraj and John Cioffi,\\
Electrical Engineering Department, Stanford University, CA, USA\\
$^*$Electrical and Computer Engineering Department, University of
Houston, Houston, USA\vspace{-7mm}}

\maketitle

\begin{abstract}

Treating the interference as noise in the $n-$user interference
channel, the paper describes a novel approach to the rates region,
composed by the time-sharing convex hull of $2^n-1$ corner points
achieved through On/Off binary power control. The resulting rates
region is denoted {\em crystallized} rates region.
%This paper focuses on utility maximization for the $n-$user
%interference channel using game theory concepts.
By treating the interference as noise, the $n-$user rates region
frontiers has been found in the literature to be the convex hull of
$n$ hyper-surfaces. The rates region bounded by these hyper-surfaces
is not necessarily convex, and thereby a convex hull operation is
imposed through the strategy of time-sharing. This paper simplifies
this rates region in the $n-$dimensional space by having only an
On/Off binary power control. This consequently leads to $2^n-1$ corner
points situated within the rates region. A time-sharing convex hull is imposed onto those corner
points, forming the crystallized rates region. The paper focuses on
game theoretic concepts to achieve that crystallized convex hull via
correlated equilibrium. In game theory, the correlated equilibrium
set is convex, and it consists of the time-sharing mixed strategies
of the Nash equilibriums. In addition, the paper considers a
mechanism design approach to carefully design a utility function,
particularly the Vickrey-Clarke-Groves auction utility, where the
solution point is situated on the correlated equilibrium set.
%In Mechanism design, we
%focus on employing the Vickrey-Clarke-Groves auction, which has the
%social optimum property.
Finally, the paper proposes a self learning algorithm, namely the regret-matching algorithm,
that converges to the solution point on the correlated equilibrium set
in a distributed fashion.

%Note: the crystal part is too heavy than the game part.

%Here it is not the game between different users. Instead, users
%belong to the same authority and have similar objective function.
%The key is to find the distributive optimization over interference
%channel.
%
%Note: non-regret algorithm can surely find the convex hull of the
%Nash equilibrium. Here how to make sure that any point on the
%boundary of the capacity region is a Nash equilibrium. Careful
%utility function design is needed.

\end{abstract}

% ------------------------------ SECTION ----------------------------------
\section{Introduction}\label{sec:intro}

Wireless systems are becoming increasingly interference limited
rather than noise limited, attributed to the fact that the cells are decreasing in size and the number of users within a cell is increasing.
 Mitigating  the
impact of interference between transmit-receive pairs
 is of great importance in order
to achieve higher data rates. Describing the complete capacity region of the
interference channel remains an open problem in information theory~\cite{Ahlswede,Carleial,Costa,Han-Kobayashi,Gans}. For very
strong interference, successive cancellation schemes have to be
applied, while in the weak interference regime, treating the
interference as additive noise is optimal to within one bit ~\cite{Tse, Veeravalli,
Kramer, Khandani}. Treating the interference as noise, the $n-$user achievable rates region has
been found in \cite{Charafeddine} to be the convex hull of $n$
hyper-surfaces.  The rates region bounded by
these hyper-surfaces is not necessarily convex, and hence a convex
hull operation is imposed through the strategy of time-sharing.

This paper adopts a novel approach into simplifying this rates
region in the $n-$dimensional space by having only On/Off
binary power control. Limiting each of the $n$ transmitters to a transmit power of either $0$ or $P_{\max}$, this consequently leads to $2^n-1$ corner points within the rates region. And by forming a convex hull through time-sharing between those corner points, it thereby leads to what we denote a crystallized rates region.

Utility maximization using  game-theoretic
techniques has recently received significant attention
\cite{Larsson,Larsson2,Cioffi,JSAC,Cioffi2}. Most of the existing game
theoretic works are based on the concept of Nash equilibrium
\cite{gamebook}. However, the Nash equilibrium investigates the
individual payoff and might not be system efficient, i.e. the
performance of the game outcome could still be improved. In 2005, Nobel
Prize was awarded to Robert J. Aumann for his contribution of
proposing the concept of correlated equilibrium \cite{Aumann74}.
Unlike Nash equilibrium in which each user only considers its
own strategy, correlated equilibrium achieves better performance by
allowing each user to consider the joint distribution of the users'
actions. In other words, each user needs to consider the others'
behaviors to see if there are mutual benefits to explore
\cite{Altman,Altman2,Han1}. Likewise, mechanism design (including
auction theory \cite{Auction}) is a subfield of game theory
that studies how to design the game rule in order to achieve good
overall system performance \cite{papadimitriou,Hart_Mas-Colell00}.
Mechanism design has drawn recently a great attention in the research
community, especially after another Nobel Prize in 2007.

The paper presents three contributions with the following structure:
\begin{enumerate}

\item Section \ref{sec:Crystallized} introduces the concept of
crystallized rates region with On/Off power control.

\item Section \ref{sec:CE} applies the game theoretic concept of
correlated equilibrium (CE) to the rates region problem. The CE
exhibits the property of forming a convex set around the $2^n-1$
corner points, hence fitting suitably in the crystallized rates
region formulation.

\item Using mechanism design, Section \ref{sec:mechanismDesign}
presents an example in applying these two concepts for the $2-$user
channel and formulates the Vickrey-Clarke-Groves auction utility. To
find the solution point distributively, the regret matching learning
algorithm is employed by virtue of its property of converging to the
correlated equilibrium set.
\end{enumerate}
Section \ref{sec:simulation} demonstrates the ideas through
simulation, and Section \ref{sec:conclusion} draws the conclusions.

%With closer Base Stations in cellular communication, and with
%closer transceiver nodes in ad hoc networks, interference is
%becoming the limiting factor to achieve higher rates.
%Traditionally for noise-limited regime, power control is the main
%concern for utility optimization. Therefore, there should be a
%paradigm shift from focusing on power control to focusing on
%time-sharing for the interference-limited regimes. In approaching
%the problem of the $n-$user interference channel, we apply the
%scheme of Binary Power Control Time-Sharing (BPCTS) . Such
%approach results in the concept of a crystallized rates region (see
%Fig ??). Such scheme would achieve the optimal frontier whenever
%the system becomes increasingly interference limited. We discuss
%this in more details in Section  ??.

%Our contributions in this paper are:
%\begin{itemize}
%\item We apply the concept BPCTS to form a crystallized rates region, which renders utility maximization more tractable. Such approach is of increased importance in the shift towards interference-limited communication.
%\item We use C.E. to approach the problem due to its convexity nature of the solution set.
%\item We focus on the VCG auction which has the social optimum property of truth revealing [?], Non-regret learning is used to solve the problem in a distributive manner.
%\item We show through simulation the gradual tendency towards time-sharing as interference gain increases. Two users and 3-user scenarios are evaluated.
%\end{itemize}

\vspace{-2mm}

% ------------------------------ SECTION ----------------------------------
\section{Crystallized Rates Region}\label{sec:Crystallized}

\subsection{System Model for $2-$user Interference Channel}

\begin{figure}[h]
\centering
\includegraphics[width=.45\textwidth]{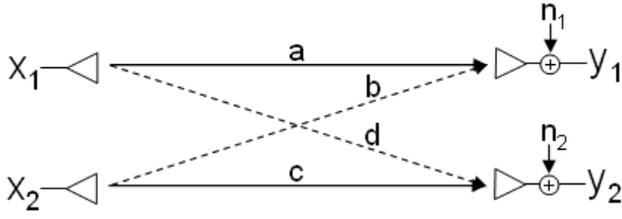}
\caption{$2-$user interference channel}
\label{fig:2usersSys}\vspace{-5mm}
\end{figure}

A $2-$user interference channel is illustrated in Fig.~\ref{fig:2usersSys}. User $i$ transmits its signal $X_i$ to the
receiver $Y_i$. The receiver front end has additive thermal noise
$n_i$ of variance $\sigma_n^2$. There is no cooperation at the transmit, nor at the receive side.
The channel is flat fading. For brevity $a$, $b$, $c$, and $d$
represent the channel power gain {\em normalized} by the noise
variance. Explicitly, $a=|g_{1,1}|^2/\sigma_n^2$,
$b=|g_{2,1}|^2/\sigma_n^2$, $c=|g_{2,2}|^2/\sigma_n^2$, and
$d=|g_{1,2}|^2/\sigma_n^2$, where $g_{i,j}$ is the channel gain from
the $i^{th}$ transmitter to the $j^{th}$ receiver. User $i$
transmits with power $P_i$, and it has a maximum power constraint of
$P_{\max}$. %The maximum power constraint follows from the practical
%communication systems implementation, where it is desired to keep
%the transmit power amplifier within its operational linear range.

In an effort to keep the complexity of the receivers fairly simple,
the interference is treated as noise. Such case is encountered in sensor
networks and in cellular communication where it is desired to have
low power-consuming and correspondingly low complexity receivers. Therefore, with the power vector ${\bf P} = [P_1, P_2]^T$,
treating the interference as noise the achievable rates for the
$2-$user interference channel are:
\begin{align}
\begin{array}{l}
R_1({\bf P})=\log_2\left(1+\frac{\displaystyle aP_1}{\displaystyle 1 +
bP_2}\right);\\ R_2({\bf P})=\log_2\left(1+\frac{\displaystyle
cP_2}{\displaystyle 1 + dP_1}\right).
\end{array}%\nonumber
\label{ratesEquations}
\end{align}

\subsection{The Achievable Rates Region Treating Interference as Noise}
In \cite{Charafeddine} the achievable rates region for the general $n-$user channel, treating the interference as noise, is found to be the convex hull of the union of $n$ hyper-surfaces. Each hyper-surface is characterized by holding one of the transmitter constant at a full
power while the other transmitters sweep all their power range, hence
forming a hyper-surface as a result. There are $n$ transmitters, resulting in $n$ hyper-surfaces, onto which the convex hull operation is performed.

The convexity or concavity behavior of these hyper-surface frontiers is complex. A rates region set is convex whenever it entirely encloses a straight line formed by connecting any two points within the rates region. As a result when the rates {\em region} is {\em convex}, its outerbound hyper-surface {\em frontiers} are {\em concave}, and vice versa.

\begin{figure*}[t]
\centering
\includegraphics[width=.22\textwidth]{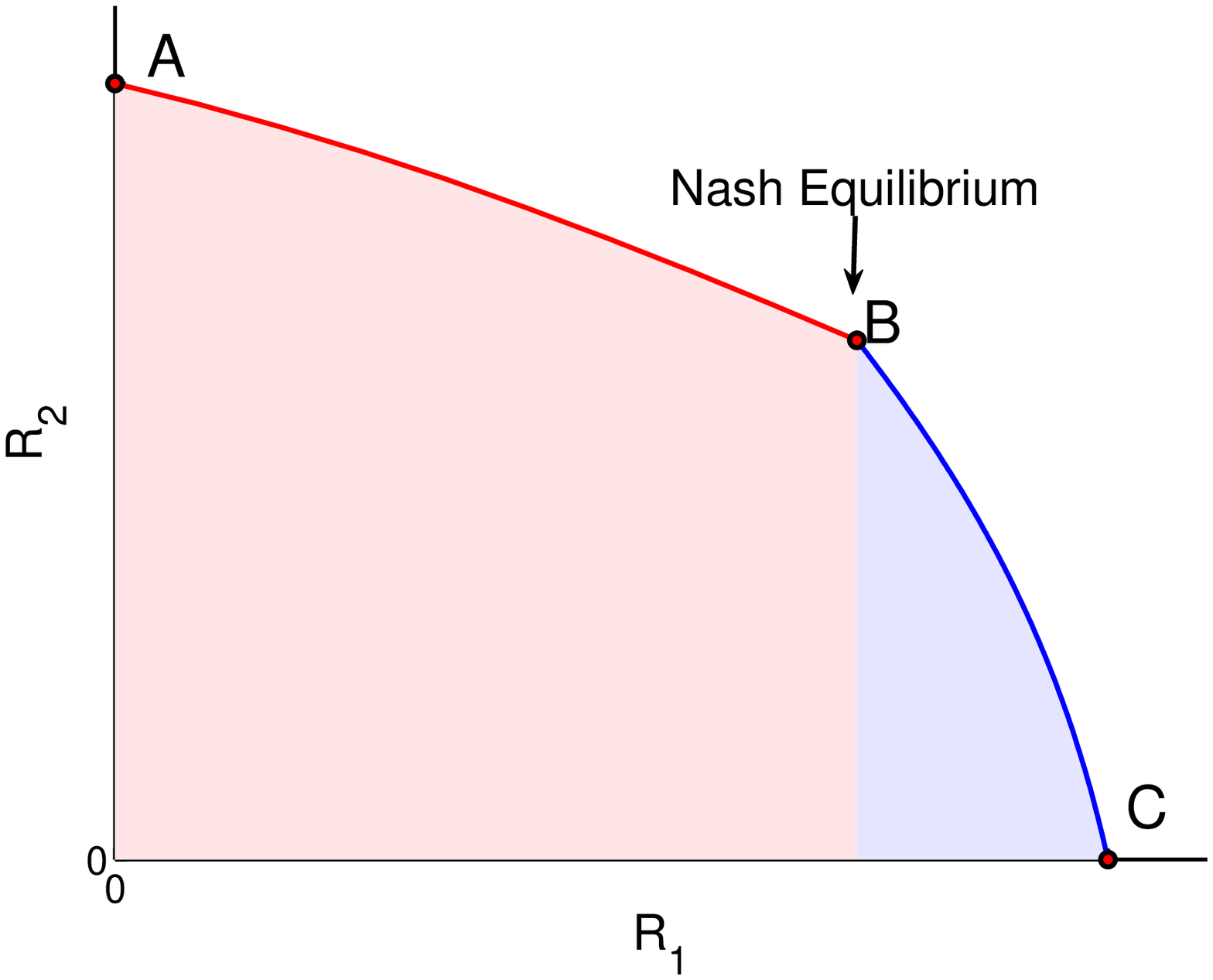}\hspace{5mm}
\includegraphics[width=.22\textwidth]{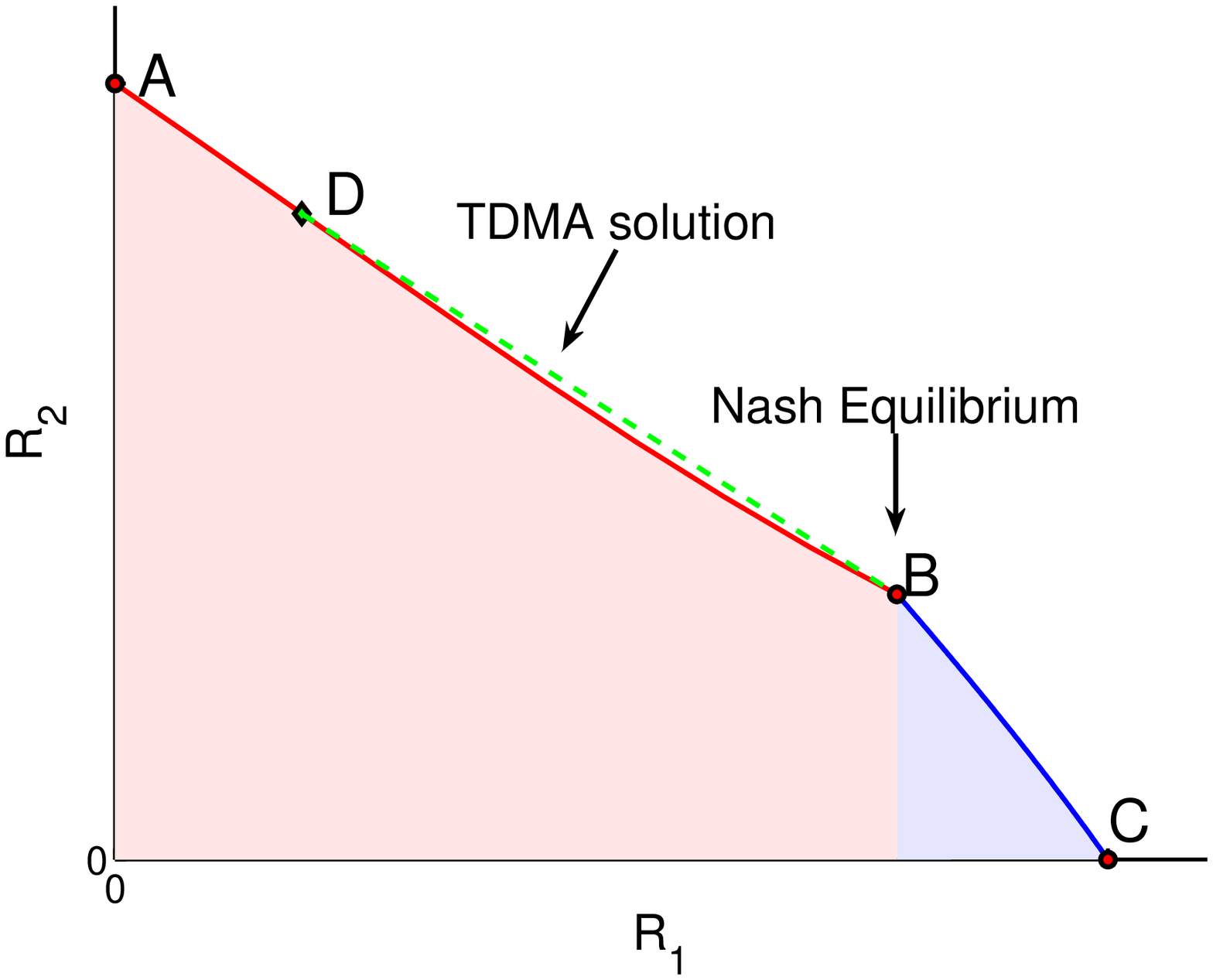}\hspace{4mm}
\includegraphics[width=.22\textwidth]{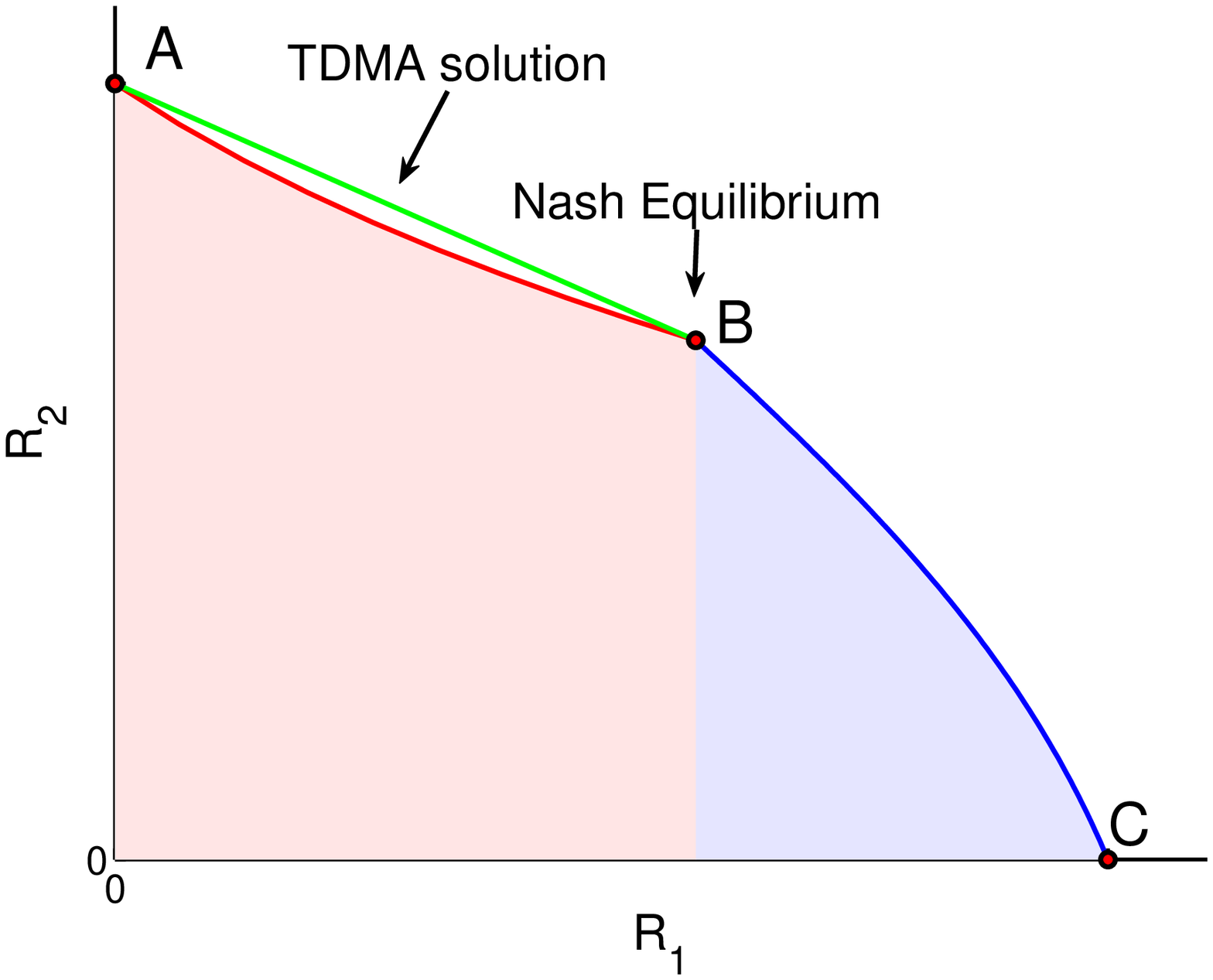}\hspace{3mm}
\includegraphics[width=.22\textwidth]{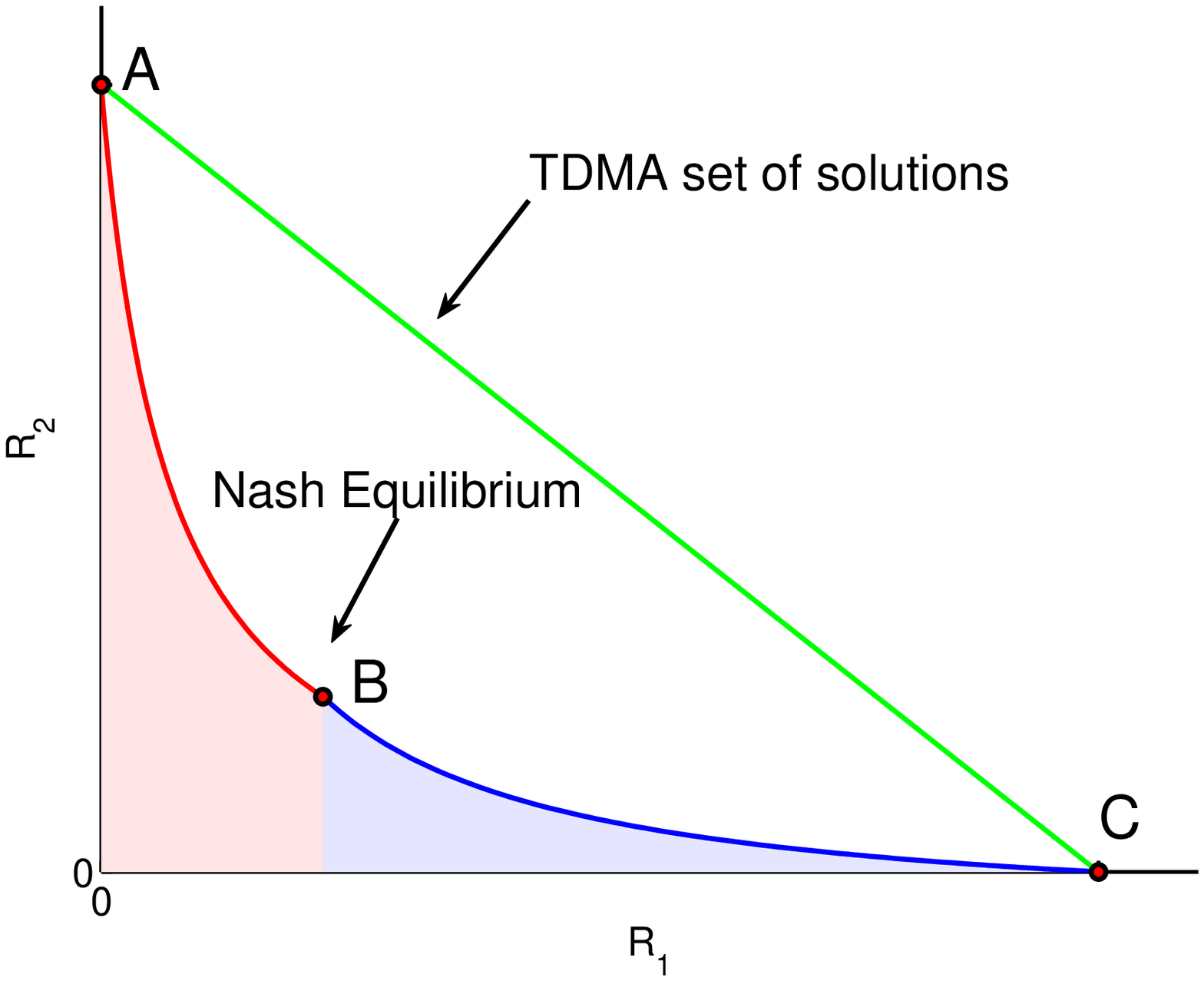}
\caption{$2-$user rates region: (a) noise-limited, concave frontiers ($\Phi_{AB}$,$\Phi_{BC}$);
(b) frontier ($\Phi_{AB}$) with inflection-point; (c) convex ($\Phi_{AB}$) and concave ($\Phi_{BC}$) frontiers; (d)
interference-limited, convex frontiers ($\Phi_{AB}$,$\Phi_{BC}$). } \label{fig:2Drateregion}
\end{figure*}

 For the $2-$user channel, see Fig.~\ref{fig:2Drateregion}, the hyper-surfaces  are the two frontiers: $\Phi_{AB}=\Phi(:,P_{\max})$, characterized by holding $P_2=P_{\max}$ and $P_1$ sweeps all its power range from $0$ to $P_{\max}$, and $\Phi_{BC}=\Phi(P_{\max},:)$ characterized by holding $P_1=P_{\max}$ and $P_2$ sweeps all its power range from $0$ to $P_{\max}$. These frontiers are referred to as {\em potential lines} given that each is characterized by holding one the transmit power arguments at a constant value, in this case $P_{\max}$, while the other power argument spans the whole power range.

These potential lines are concave in noise-limited regimes (thus enclosing a convex rates region) as in Fig.~\ref{fig:2Drateregion}-a, and they shift towards convexity as the interference increases, as in Fig.~\ref{fig:2Drateregion}-d. In cases with moderate interference levels, they can exhibit non-stationary inflection point, as at point D in Fig.\ref{fig:2Drateregion}-b.

\subsection{Crystallized Rates Region}
\label{subsec:crystallized}

The crystallized rates region approach approximates the achievable
rates region formed by the potential lines $\Phi_{AB}$ and
$\Phi_{BC}$ into the convex time-sharing hull of the straight lines
connecting points A, B, and C. Denoting $\Phi(P_1,P_2)$ the point in the
rates region achieved when user $1$ transmits at $P_1$ and user $2$
transmits at $P_2$ in Eq.~(\ref{ratesEquations}): point A is
$\Phi(0,P_{\max})$ where only user 2 transmits at full power and
user 1 is silent, point B is $\Phi(P_{\max},P_{\max})$ where both
users transmit simultaneously at full power, and point C is
$\Phi(P_{\max},0)$ where user 1 transmits at full power and user 2
is silent. Hence, we refer by binary power control such mechanism of
operation in producing points A, B, and C. We denote such points
that are formed by binary power control as the {\em corner} points of the
rates region. In the $2-$user interference channel, there exist $3$
corner points, similarly in the $n-$user case there exist $2^n-1$
corner points.

\begin{figure*}[t]
\centering
\includegraphics[width=.35\textwidth]{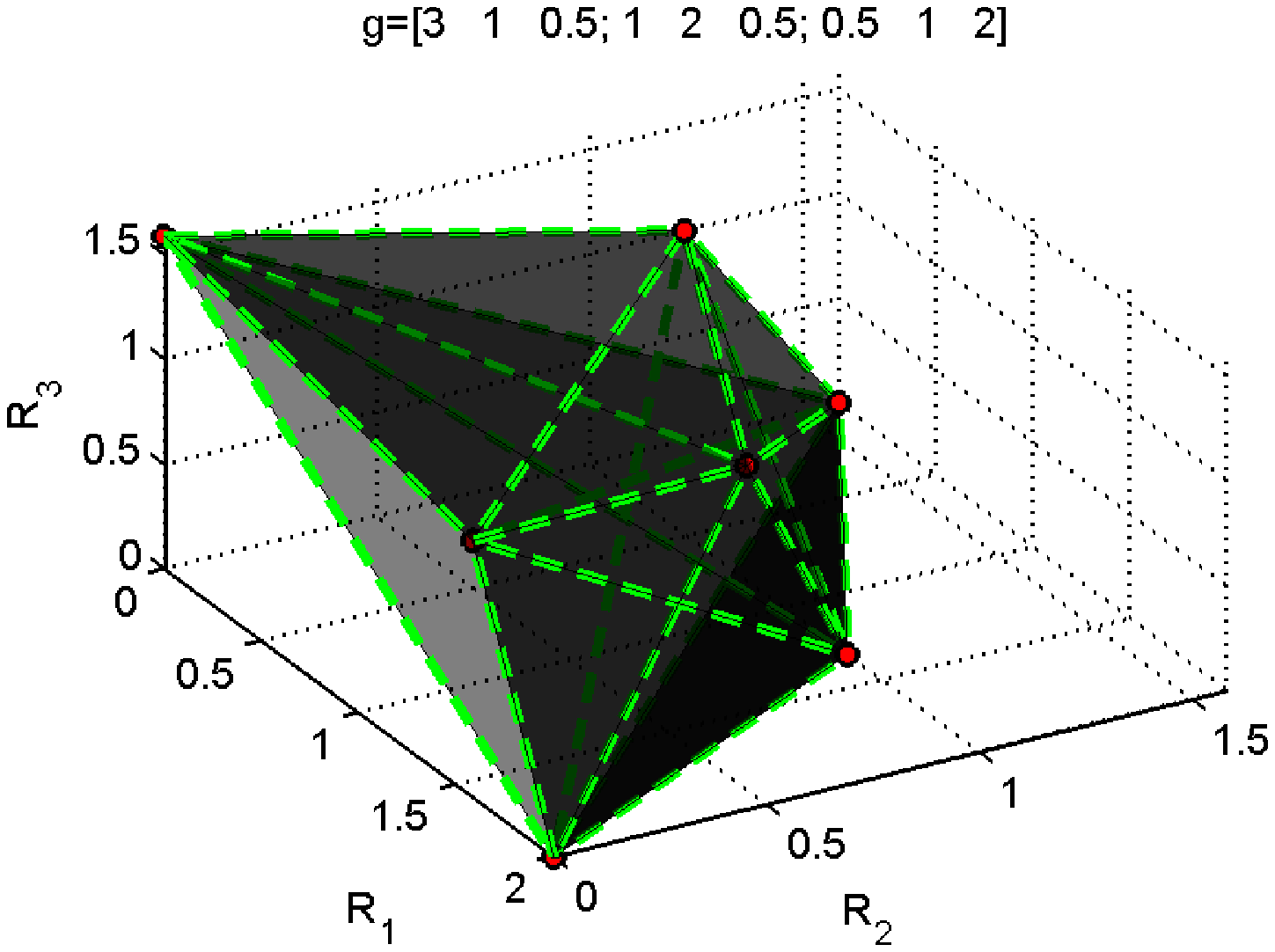}
\includegraphics[width=.35\textwidth]{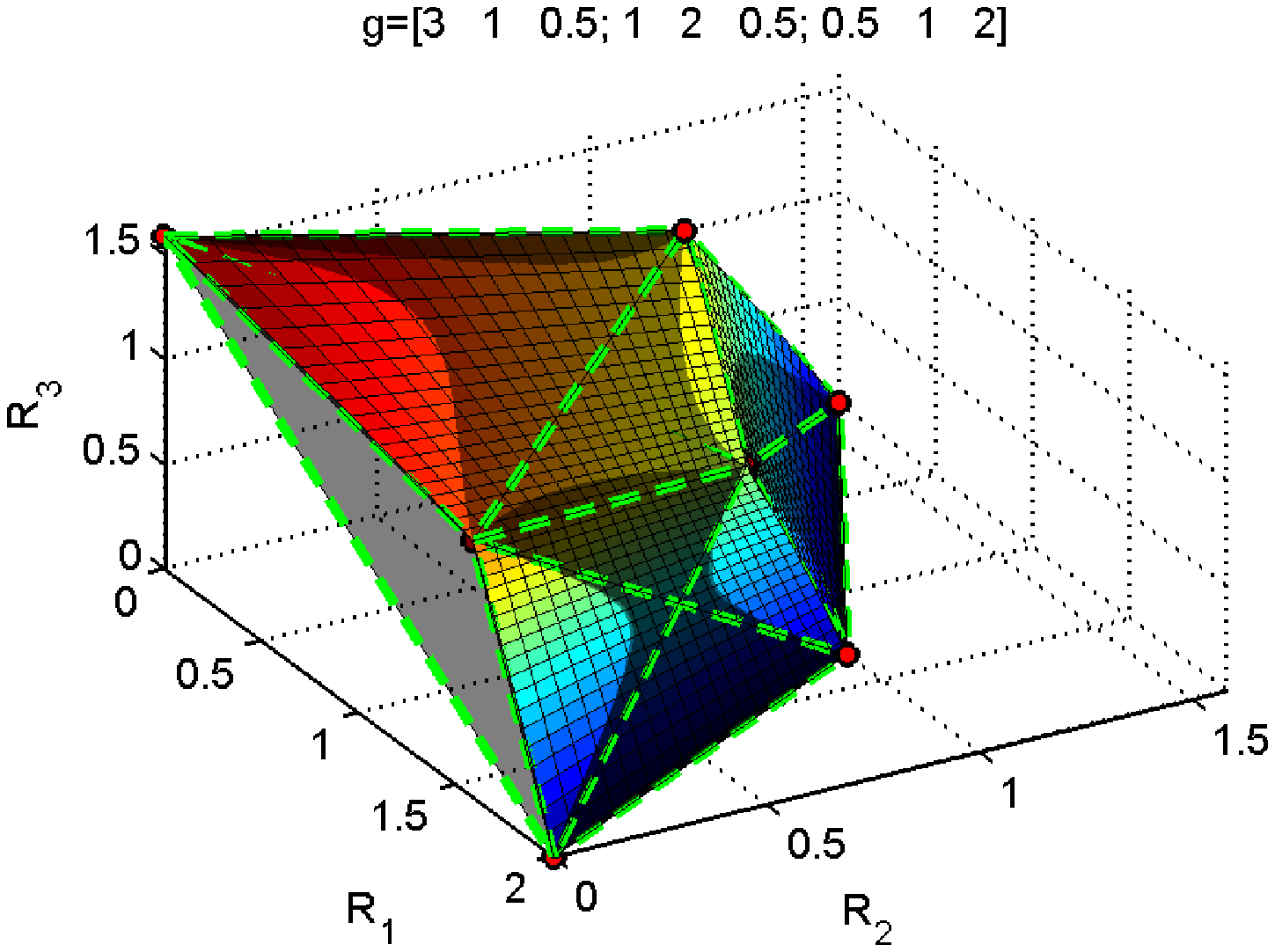}
%\hspace{1cm}
%\includegraphics[width=.3\textwidth]{g_3D_interfLimited.eps}
\caption{$3-$user crystallized rates region: (a) time-sharing crystallized hull,(b) crystallized hull overlaid on top of the rates region} \label{fig:3users}
\end{figure*}

Therefore, this paper simplifies the analysis of the rates region in
the $n-$dimensional space to just focus on finding the convex
time-sharing hull onto the $2^n-1$ corner points, forming the 
crystallized rates region. In the $2-$user dimension, these are
straight time-sharing lines connecting two points; in the $3-$user
dimension, these are a set of polygon surfaces each connecting three
points, see Fig. \ref{fig:3users}.  %Such convex time-sharing hull
%can be attained via the inherently convex Correlated Equilibrium set
%using the appropriate tools in game theory.

%Such convex hull
%operation comes readily using the Correlated Equilibrium concept
%in game theory; which is the focus of the following section.

%\tableofcontents

%\subsection{Problem Formulation}
%
%For system point of view, we want to optimize the social optimum
%as
%\begin{equation}
%\max _{p_{i,j}} \sum_{i=1}^2 \sum_{j=1}^2 F(R_1, R_2) p_{i,j}
%\end{equation}
%where $F(R_1,R_2)$ can be weighted sum rate, maximin rate, or
%proportional fair rate, etc. In this paper, we use sum rate for
%simplicity. Other definitions of social optimum can be employed in
%a similar way.

\subsection{System Time-sharing Coefficients and Rates Equations}
\label{subsec:thetas}
Instead of a power control problem in finding $P_i$, the problem becomes finding the appropriate time-sharing coefficients of the $2^n-1$ corner points. For the $2-$user case, let {\boldmath $\theta$}$=[\theta_1, \theta_2, \theta_3]^T$, $\sum_i\theta_i=1$, denote the {\em system} time-sharing coefficients vector of the respective corner points $\Phi(P_{\max},0)$ (user 1 transmitting only with a time-sharing coefficient $\theta_1$), $\Phi(0,P_{\max})$ (user 2 transmitting only with a time-sharing coefficient $\theta_2$), and $\Phi(P_{\max}, P_{\max})$ (both users transmitting with  a time-sharing coefficient $\theta_3$). The reason {\boldmath $\theta$} is labeled a {\em system} time-sharing coefficients vector is to emphasize the combinatorial element in constructing the corner points, where the cardinality of $|${\boldmath $\theta$}$|=2^n-1$.

Then for $2-$user case, in contrast with Eq.~(\ref{ratesEquations}), the new crystallized rates equations for $R_1$ and $R_2$ are:
\begin{align}%\hspace{-.25cm}
\begin{array}{l}
R_1(\mbox{\boldmath $\theta$})=\theta_1\log_2(1+aP_{\max})+\theta_3\log_2\left(1+\frac{\displaystyle aP_{\max}}{\displaystyle 1+bP_{\max}}\right)\\
R_2(\mbox{\boldmath$\theta$})=\theta_2\log_2(1+cP_{\max})+\theta_3\log_2\left(1+\frac{\displaystyle cP_{\max}}{\displaystyle 1+dP_{\max}}\right)
\end{array}
\label{crystallizedRates}
\end{align}

Any solution point on the crystallized frontier would lie somewhere on the
time-sharing line connecting two points for the $2-$user case; and similarly for the
$3-$user case, the solution point lies somewhere on a time-sharing
plane connecting three points, then by deduction we obtain the
following corollary:
\begin{corollary}
\label{cor:max_n_nonzero}
The system time-sharing vector {\boldmath $\theta$}, for any solution point on the
$n-$user crystallized rates region, has at maximum $n$ nonzero coefficients out of its $2^n-1$ elements.
\end{corollary}

%{\em Corollary I:} As in the $2-$user case, the solution point lies
%on a time-sharing line connecting two points, and similarly in the
%$3-$user case, the solution point lies on a time-sharing plane
%connecting three points, then any solution point in the $n-$user
%dimension has at maximum $n$ nonzero coefficients inside the
%($2^n-1$)-elements vector {\boldmath $\theta$}.

\subsection{Evaluation of Crystallization}

Examining the crystallized rates region in more details for the
$2-$user interference channel, we evaluate the area of the rates
region bounded by the potential lines $\Phi_{AB}$ and $\Phi_{BC}$
achieved through power control, and the area of the rates region
formed by time-sharing points A, B, and C. In effect, we are
evaluating how much gain or loss results from completely replacing
the traditional power control scheme (see
Eq.~(\ref{ratesEquations})) with the time-sharing scheme between the
corner points (see Eq.~(\ref{crystallizedRates})). For this purpose
we consider the symmetric channel, where $a=1$, and we increase the
interference $b$ to vary the signal to interference ratio $SIR =
a/b$ from $20$dB to $-20$dB.
\begin{figure}[h]
\centering
\includegraphics[width=.3\textwidth]{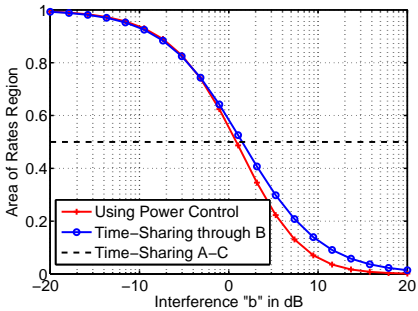}
\caption{Area of the rates region achieved through power control or
through time-sharing versus interference}
\label{fig:RRAreas}\vspace{-5mm}
\end{figure}
The value of the area bounded by the power control potential lines
is plotted in Fig.\ref{fig:RRAreas} together with the value of the
area bounded by the time-sharing scheme through the point B ( formed
by the time-sharing lines A-B and B-C). In addition, for reference,
the area confined by the time-sharing line A-C is plotted, which
does not depend on the SIR.

For weak interference, or equivalently noise-limited regime, point B is used in constructing the crystallized region. As the interference increases beyond a certain threshold level, time-sharing through point B becomes suboptimal, and time-sharing A-C becomes optimal. The exact switching point from power control to time-sharing has been found in \cite{Charafeddine}.  In Fig.~\ref{fig:RRAreas}, this happens at the intersection of the blue line (with circle markers) and the A-C dotted line. As indicated in Fig.~\ref{fig:RRAreas}, there is no significant loss in the rates region area if time-sharing is used universally instead of traditional power control, in fact in some cases time-sharing offers considerable gain. Specifically, whenever the potential lines exhibit concavity, time-sharing loses to power control; whenever the potential lines exhibit convexity, time-sharing gains over power control.  Different values of $a$ also lead to the same conclusion.

\begin{figure}[h]
\centering
\includegraphics[width=.3\textwidth]{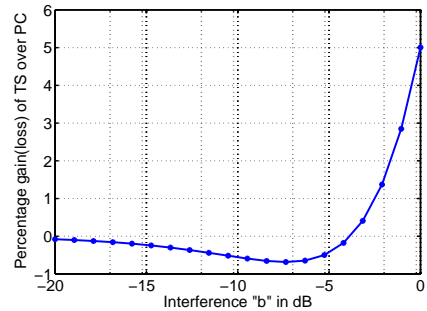}
\caption{Gain (or loss) percentage from using time-sharing through point B over power control}
\label{fig:RRAreasPerct}\vspace{-5mm}
\end{figure}

In Fig.~\ref{fig:RRAreasPerct} the percentage of the rates region
gain (or loss)   in using the time-sharing scheme (through point B)
over the power control scheme is plotted for the same symmetric
channel examined in Fig.~\ref{fig:RRAreas}. The loss does not exceed
$1\%$, and the time-sharing strategy is therefore quite attractive.
For illustration purposes, note that the x-axis in
Fig.~\ref{fig:RRAreasPerct} was chosen to span the interference
range of $-20$dB to $0$dB; whereas in Fig.~\ref{fig:RRAreas} the
x-axis interference range was from $-20$dB to $20$dB. If we were to
plot the x-axis in Fig.~\ref{fig:RRAreasPerct} up to $20$dB instead
of the $0$dB, the percentage gain would have reached on the y-axis
up to $800\%$.  Note that for high interference time-sharing through point B is suboptimal
and time-sharing A-C is optimal, so the gain over power control is even larger.

\section{Correlated Equilibrium for Crystallized Interference Channel}\label{sec:CE}

The crystallized rates region offers a good alternative to form the rates region of the $n-$user interference channel with marginal loss, and sometimes significant gain (especially for interference-limited regimes). Therefore the problem revolves around finding the convex hull over the set of polygons connecting the $2^n-1$ corner points. One technique
explored to achieve the convex hull is through the concept of correlated equilibrium in game theory.

\subsection{Correlated Equilibrium}
\label{subsec:CE}

Every user $i$ has a transmit strategy $\alpha_i$ of either $0$ or $P_{\max}$. $U_i$ is the utility of user $i$. Nash equilibrium is a
well-known concept to analyze the outcome of a game which states that in the equilibrium every
user will select a utility-maximizing strategy given the
strategies of every other user.
\begin{definition}
Nash equilibrium achieving strategy $ \alpha_i^*$ of user $i$ is defined as:
\begin{align}
   U_i( {\alpha}^*_i,  \mbox{\boldmath $\alpha$}_{-i})\geq U_i(
    {\alpha}_i, \mbox{\boldmath $\alpha$}_{-i}),\ \forall i,\ \forall
     {\alpha}_i \in \Omega_i
\end{align}
where ${\alpha}_i$ is any possible strategy of user $i$, \mbox{\boldmath $\alpha$}$_{-i}$ is the strategy vector of all other users except user $i$, and
$\Omega_i$ is the strategy space $\{0,P_{\max}\}$. In other words, given the
other users' actions, no user can increase its utility alone by
changing its own action.
\end{definition}

%If a user will follow an action in every possible attainable
%situation in a game, the action is called pure strategy. In the
%case of mixed strategies, the user will follow a probability
%distribution over different possible actions. For example, in Fig.
%\ref{fig:2Drateregion} (a), Point B is a pure Nash equilibrium;
%and in Fig. \ref{fig:2Drateregion} (d), any time-sharing point
%between Point A and Point C is a mixed strategy.

%In Table
%\ref{table:Reward}, we illustrate an example of two users with
%different actions. In Table \ref{table:Reward} (a), we list the
%utility values for two users taking actions $0$ and $1$. We can
%see that when both of the two users take action $0$, they have the
%best overall benefit. We can view this action as a cooperative
%action. But if any user plays more aggressively using action $1$
%while the other still plays action $0$, the aggressive user has a
%better utility, but the other user has a lower utility and the
%overall benefit is reduced. However, if both users play
%aggressively using action $1$, both of them obtain very low
%utilities. In Table \ref{table:Reward} (b), we show two Nash
%equilibria, where one of the user dominates the other. The
%dominating user has the utility of $6$ and the dominated user has
%the utility of $3$, which is unfair. In Table \ref{table:Reward}
%(c), we show the mixed Nash equilibrium where two users have the
%probability $0.75$ for action $0$ and $0.25$ for action $1$,
%respectively. The utility for each user is $4.5$.
%\vspace{-2mm}

%\subsection{Correlated Equilibrium}

Next the concept of the correlated equilibrium is studied. It is
 more general than the Nash equilibrium and it was first proposed in
\cite{Aumann74}. The idea is that a strategy profile is chosen
according to the joint distribution instead of the marginal
distribution of users strategies. When converging to the recommended strategy, it is to the users'
best interests to conform to this strategy. The distribution is
called the correlated equilibrium, which is defined as:
\begin{definition}
A probability distribution $p$ is a correlated equilibrium of a
game, if and only if, for all $i$, ${\alpha}_i\in \Omega_i$, and
$\mbox{\boldmath $\alpha$}_{-i}\in \Omega_{-i}$,
\begin{equation}\label{eqn:CE1}
\sum_{\mbox{\boldmath $\alpha$}_{-i}\in \Omega_{-i}}
p({\alpha}_{i}^*,\mbox{\boldmath $\alpha$}_{-i})[U_i({\alpha}^*_i, \mbox{\boldmath $\alpha$}_{-i})-U_i({\alpha}_i, \mbox{\boldmath $\alpha$}_{-i})]\geq0, \forall \alpha_i\in \Omega_i.
\end{equation}
\end{definition}
$\Omega_{-i}$ denotes the strategy space of all the users other than user $i$. As every user $j$, $j\neq i$, has a possible $0$ or $P_{\max}$ strategy choice, then the cardinality of $\Omega_{-i}$ is $|\Omega_{-i}|=2^{(n-1)}$. Therefore the summation in Eq.~(\ref{eqn:CE1}) have $2^{n-1}$ summation terms. The summation over $\mbox{\boldmath $\alpha$}_{-i}$ generates the marginal expectation. The inequality (\ref{eqn:CE1}) means that when the recommendation to user $i$ is to choose action ${\alpha}_i^*$, then choosing action ${\alpha}_i$ instead of ${\alpha}_i^*$ cannot result in a higher expected payoff to user $i$. It is worth to point out that the probability distribution $p$ is a joint point mass function (pmf) of the different combinations of the $n$ users strategies. Therefore, $p$ is the joint pmf of the resulting $2^n$ {\em system} strategy points. Discounting the trivial system strategy of all the users transmitting at $0$, there exist $2^n-1$ system strategy points that we wish to find their pmfs.

\subsection{CE in the Context of the Crystallized Rates Region}
\label{subsec:CEinCrys}

Revisiting Subsection \ref{subsec:thetas}, the $2^n-1$ point mass
functions that we want to find are the system time-sharing
coefficients $\theta_k$, $k=1,\ldots,2^n-1$. We can index those
$2^n-1$ pmfs to the corresponding $\theta_k$ in any bijective
one-to-one mapping. Index $k$ can denote the base-$2$ representation
of the binary users' strategies (starting with user $1$'s binary
action as the least significant bit). For example, let
$\alpha_i^{(1)}$ denotes that user $i$ transmits with
$\alpha_i=P_{\max}$, and $\alpha_i^{(0)}$ denotes that user $i$ is
silent with $\alpha_i=0$. In Subsection \ref{subsec:thetas},
$\theta_1$ was mapped to user $1$ transmitting, equivalently
$\theta_1=p(\alpha_1^{(1)},\alpha_2^{(0)})=p_{\Phi(P_{\max},0)}$;
where we defined explicitly $p_{\Phi(P_{\max}, 0)}$ as the point
mass function of the point $\Phi(P_{\max}, 0)$. And similarly
$\theta_3$ was mapped to both users transmitting,  $\theta_3=
p(\alpha_1^{(1)},\alpha_2^{(1)})=p_{\Phi(P_{\max}, P_{\max})}$.
Morever, by definition, $\sum_{\mbox{\boldmath
$\alpha$}}p(\mbox{\boldmath
$\alpha$})=\sum_{k=1}^{(2^n-1)}\theta_k=1$, and as discussed in
Corollary I, the solution point possesses at most $n$ nonzero pmfs
in the joint distribution $p$.

%A remark here: please note that in Eq (\ref{eqn:CE1}), if the joint probability is flat, i.e. homogenous for every $i$, then Eq (\ref{eqn:CE1}) simplifies and the joint probability factor $p({\alpha}_{i},{\alpha}_{-i})$ drops . The NE problem is $\forall i$, whereas the CE problem is cumulative (it  is the $summation$) instead of $\forall -i$.

The correlated equilibriums set is nonempty,
closed and convex in every finite game \cite{Aumann74}. In fact,
every Nash equilibrium  and mixed (i.e. time-sharing) strategy of Nash equilibriums are
within the correlated equilibrium set, and the Nash equilibrium
correspond to the special case where $p(\mbox{\boldmath $\alpha$})$
is a product of each individual user's probability for different
actions, i.e., the play of the different users is independent
\cite{Aumann74,Hart_Mas-Colell00}.

% In general, if global information is
%known the correlated equilibrium
%can be calculated by linear programming.

%In our study of the interference channel if $U_i=R_i,\forall i$,
%Nash equilibrium is achieved when $\alpha_i=1,\forall i$, which
%corresponds to point $B$ in Fig. \ref{fig:2Drateregion} . The
%correlated equilibrium only consists of one point which is the Nash
%equilibrium. However, except for Fig. \ref{fig:2Drateregion} (a),
%the optimal solution from sum rates point of view might not
%have all $\alpha_i$ to be $1$. For example, in Fig.
%\ref{fig:2Drateregion} (d), the optimal solution appears on the line
%between point $A$ and $C$. So the question is how to let the corner
%points (such as $A$ and $C$ in Fig. \ref{fig:2Drateregion} (d)) to
%be the Nash equilibria, and let hyper-surfaces (such as the line
%between $A$ and $C$ in Fig. \ref{fig:2Drateregion} (d)) consisting
%those corner points to be the correlated equilibrium set, if we can
%design the game carefully. We will answer this question the following Section.

%If we can let the corner points of the hyper-surfaces to be the Nash
%equilibriums for the time sharing binary power control case, we can
%employ the concepts of the correlated equilibrium.

\section{Mechanism Design and Learning Algorithm}
\label{sec:mechanismDesign}

There are two major challenges to implement correlated equilibrium
for rate optimization over the interference channel. First, to
ensure the system converges to the desired point (such as
time-sharing between A-C instead of going through point B in Fig.
\ref{fig:2Drateregion} (d)). As an example, we considered an auction utility function from mechanism design.
Second, to achieve the equilibrium,
a distributive solution is desirable, where we propose the
self-learning regret matching algorithm.

\subsection{Mechanism Designed Utility}

%In this section, based on the idea of mechanism design, we design a
%utility such that the corners of the hyper-surfaces feasible region
%are the Nash equilibria. By doing this, if we can find the set of
%the correlated equilibrium, we also find the convex hull of the
%hyper-surfaces feasible region. Then we propose a distributed
%learning algorithm to achieve the correlated equilibrium set.

%<Zhu21Sept> note: Zhu commented this table! </Zhu21Sept>
%\begin{table}[ht]
 % \caption{Two-user game modified utility table}\label{table:modified chick}
 % \centering\normalsize{
 % \vspace{3mm}
 % \begin{tabular}{|c|c|c|}
 %  \hline
  %  & $P_2=0$ & $P_2=P_{\max}$ \\
  %  \hline
  %  $P_1=0$ & ($0$,$0$) &  ($0$,$\log_2(1+cP_{\max})$) \\
  %  \hline
  %  $P_1=P_{\max}$ & ($\log_2(1+aP_{\max})$,$0$)&
  %  ($U'_{1}$,$U'_{2}$)\\
%    \hline
 % \end{tabular}
% }\vspace{-3mm}
%\end{table}

One important mechanism design is the Vickrey-Clarke-Groves (VCG)
auction \cite{Auction} which imposes cost to resolve the conflicts
between users. Using the basic idea of VCG, where we want to
maximize $U_i, \forall i$, the user utility $U_i$ is designed to be
the rate $R_i$  minus a payment cost function $\zeta_i$ as
\begin{equation}\label{new_utility}
U_i\triangleq R_i-\zeta_i.
\end{equation}
%Define $\alpha^{\ast}_{-i}$ as the social optimal time-sharing
%strategy of all other users without user $i$, i.e.,
%\begin{align}
%\alpha^{\ast}_{-i}=\arg\max_{\alpha_{-i}}\sum_{j}U_{j}\left(
%\alpha_{j}\right)  ,\ \alpha_{i}=0,
%\label{eq_total_rate_max_without_i}%
%\end{align}
%Define $\alpha^{\ast}$ as the social optimal strategy with all
%users:
%\begin{align}
%\alpha^{\ast}=\arg\max_{\alpha}\sum_{j}U_{j}\left(
%\alpha_{j}\right).
%\end{align}
The payment cost of user $i$ is expressed as the performance loss of all other users
due to the inclusion of user $i$, explicitly:
\begin{equation}
\zeta_{i}(\mbox{\boldmath $\alpha$})=\sum_{j\neq
i}R_{j}\left(\mbox{\boldmath $\alpha$}_{-i}\right) -\sum_{j\neq
i}R_{j}\left( \alpha_i\right).
\end{equation}Hence if $\alpha_i$ is $0$ for user $i$, it is equivalent to user $i$ being absent, consequently the cost $\zeta_i= 0$ whenever $\alpha_i=0$. For the $2-$user case, focusing on $\zeta_1$ when $\alpha_1=P_{\max}$, hence: a) if $\alpha_2=0$, then $R_2=0$ and $\zeta_1=0$; b) if  $\alpha_2=P_{\max}$, then:
\begin{align}
\lefteqn{\zeta_1(\alpha_1=P_{\max},\alpha_2=P_{\max})}\nonumber \\
&
\begin{array}{ll}
& =R_2(\alpha_1=0,\alpha_2=P_{\max}) - R_2(\alpha_1=P_{\max},\alpha_2=P_{\max}) \\
& = \log_2\left(1+cP_{\max})-\log_2(1+\frac{\displaystyle cP_{\max}}{\displaystyle 1+dP_{\max}}\right) \\
& = \log_2\left(1+\frac{\displaystyle cdP_{\max}^2}{\displaystyle
1+cP_{\max}+dP_{\max}}\right).
\end{array}
\nonumber
\end{align}$\zeta_2$ follows by symmetry. As a result, the VCG utilities for the $2-$user channel are summarized in Table \ref{table:modified chick}, where
\begin{align}
\begin{array}{l}
U'_1=\log_2(1+\frac{\displaystyle aP_{\max}}{\displaystyle 1+bP_{\max}})-\log_2\left(1+\frac{\displaystyle cdP_{\max}^2}{\displaystyle 1+cP_{\max}+dP_{\max}}\right)\\
U'_2=\log_2(1+\frac{\displaystyle cP_{\max}}{\displaystyle 1+dP_{\max}})-\log_2\left(1+\frac{\displaystyle abP_{\max}^2}{\displaystyle 1+aP_{\max}+bP_{\max}}\right)\\
\end{array}\nonumber
\end{align}
\begin{table}[ht]
\caption{$2-$user VCG $\{U_1,U_2\}$ utility table}\label{table:modified chick}
\centering\normalsize{
\vspace{3mm}
\begin{tabular}{|c|c|c|}
\hline
 & $\alpha_2=0$ & $\alpha_2=P_{\max}$ \\
 \hline
 $\alpha_1=0$ & $\{0,0\}$ &  $\{0,\log_2(1+cP_{\max})\}$ \\
 \hline
 $\alpha_1=P_{\max}$ & $\{\log_2(1+aP_{\max}),0\}$&
 $\{U'_{1},U'_{2}\}$\\
  \hline
\end{tabular}
}%\vspace{-3mm}
\end{table}

Notice that each user pays the cost because of its involvement. This
cost function can be calculated and exchanged before transmission
with little signalling overhead.

\subsection{ The Regret-Matching Algorithm }
\label {subsec:LearningAlgorithm}
Finally, we exhibit the   regret-matching
algorithm \cite{Hart_Mas-Colell00} to learn in a distributive fashion how to
achieve the correlated equilibrium set in solving the VCG auction. The algorithm is named
regret-matching (no-regret) algorithm, because the stationary
solution of the learning algorithm exhibits no regret and the
probability to take an action is proportional to the ``regrets'' for
not having played the other actions. Specifically, for user $i$ there are two
distinct binary actions $\alpha_i^{(0)}$ and $\alpha_i^{(1)}$ at every time $t=T$ (where $\alpha_i^{(0)}=0$, and $\alpha_i^{(1)}=P_{\max}$).
The regret $\mathbb{R}$ of user $i$ at time $T$ for
playing action $\alpha_i^{(1)}$ instead of the other action $\alpha_i^{(0)} $
is\vspace{-1mm}
\begin{equation}\label{eqn:Regret2}
\mathbb{R}_i^T(\alpha_i^{(1)}, \alpha_i^{(0)}):=\max\{D_i^T(\alpha_i^{(1)}, \alpha_i^{(0)}),0\},\vspace{-1mm}
\end{equation}
where\vspace{-1mm}
\begin{equation}\label{eqn:Regret1}
D_i^T(\alpha_i^{(1)}, \alpha_i^{(0)})=\frac{1}{T}\sum_{t\leq T} [U^t_i(
\alpha_i^{(0)},\mbox{\boldmath $\alpha$}_{-i})-U^t_i(\alpha_i^{(1)}, \mbox{\boldmath $\alpha$}_{-i})].\vspace{-1mm}
\end{equation}
Here $U^t_i( \alpha_i^{(\cdot)}, \mbox{\boldmath $\alpha$}_{-i})$ is
the utility at time $t$ and $\mbox{\boldmath $\alpha$}_{-i}$ is
other users' actions. $D_i^T(\alpha_i^{(1)}, \alpha^{(0)}_i)$ has
the interpretation of average payoff that user $i$ would have
obtained if it had played action $\alpha_i^{(0)}$ every time in the
past instead of choosing $\alpha_i^{(1)}$. The expression
$\mathbb{R}_i^T(\alpha_i^{(1)},\alpha_i^{(0)})$ can be viewed as a
measure of the average regret. Similarly,
$\mathbb{R}_i^T(\alpha_i^{(0)}, \alpha_i^{(1)})$ represents the average regret if the alternative action has been taken.

Recalling the discussion in Subsection \ref{subsec:CEinCrys} about the mapping notation we adopted between the point mass functions $p(\alpha_1^{(\cdot)}, \alpha_2^{(\cdot)})$ and the system time-sharing coefficients ($\theta$s), then we want to find the point mass function
$p(\alpha_1^{(1)}, \alpha_2^{(0)})\equiv \theta_1$,
$p(\alpha_1^{(0)},\alpha_2^{(1)})\equiv \theta_2$, and $
p(\alpha_1^{(1)},\alpha_2^{(1)})\equiv \theta_3$. As discussed in Subsection
\ref{subsec:thetas} there exist $2^n=4$ pmfs for the $2-$user case.
For the trivial case of the origin point $\Phi(0,0)$,
$p(\alpha_1^{(0)}, \alpha_2^{(0)})=0$. We are left to obtain
$p(\alpha_1^{(1)}, \alpha_2^{(0)})$, $p(\alpha_1^{(0)},
\alpha_2^{(1)})$, and $p(\alpha_1^{(1)}, \alpha_2^{(1)})$.
Specifically\symbolfootnote[3]{{\em Note:} Solving for $n$ variables
instead of $2^n-1$ does not apply to $n \geq 3$; the $2-$user case is a
special case, as $\sum_{k=1}^{2^n-1=3}\theta_k=1$ was used to
simplify the unknowns to $2$. For $n\geq 3$, see
Fig.\ref{fig:3users}, it is not enough to find the time-sharing
strategy of {\em individual} users, as {\em ordering} needs to come
into the picture in arriving to the desired {\em system}
time-sharing coefficients. } to the $2-$user case, this simplifies
further to finding only {\em two} variables. Denoting $p_1=\theta_1$, and
$p_2=\theta_2$, then $\theta_3$ can be deduced as
$\theta_3=1-p_1-p_2$.

The details of the regret-matching algorithm is shown in Table \ref{table:regretmatching}. The probability $p_i$ is a linear function of the regret, see Eq.~(\ref{eq:regret_finding_p}).
 The algorithm has a
complexity of  $O(|\Omega_i|)=O(2)$.

%For every period $T$, let us define the relative frequency of
%users' action $\textbf r$ played till $T$ periods of time as
%following\vspace{-1mm}
%\begin{equation}\label{eqn: joint prob}
%z_T(\textbf r)=\frac{1}{T}\#\{t\leq T:\textbf  r_t=\textbf
%r\},\vspace{-1mm}
%\end{equation}
%where $\#(\cdot)$ denotes the number of times the event inside the
%bracket happens and $\textbf r_t$ is all users' action at time
%$t$.
By using the theorem in \cite{Hart_Mas-Colell00}, if every user plays according to the learning algorithm in Table
\ref{table:regretmatching}, the adaptive learning algorithm
has the property that the probability distribution found converges  on the  set of correlated equilibrium. It has been shown that the set of correlated equilibrium is nonempty,
closed and convex \cite{Aumann74}. Therefore, by using the algorithm
in Table \ref{table:regretmatching}, we can guarantee that the
algorithm converges to the set of CE as $T\rightarrow\infty$. %Next, simulation results are presented, where the learned point was found, and additional validation of the learned point in achieving the GCV was tested via exhaustive search.

%than 1 pdf solutions? \& which is best leading to optimal
%collective?.. i.e. is there concepts of local or global optimums
%here?} of game $G$, as $T\rightarrow\infty$.

\begin{table}[!t]
  \centering\normalsize{
  \caption{Regret-matching learning algorithm}\label{table:regretmatching}
  \begin{tabular}{|m{8.5cm}|}
    \hline
    Initialize arbitrarily probability for user $i$, $p_i$.\\
    \hline
     For t=2,3,4,... \\

    \ \ 1. Let $\alpha_i^{t-1}$ be the action last chosen by user $i$, and $\hat \alpha_i^{t-1}$ \\
     \ \ \ \ \ as the other action.\\
    \ \ 2. Find $D_i^{t-1}(\alpha_i^{t-1},\hat \alpha_i^{t-1})$ as in Eq.~(\ref{eqn:Regret1}).\\
     \ \ 3. Find average regret $\mathbb{R}_i^{t-1}(\alpha_i^{t-1}, \hat \alpha_i^{t-1})$ as in
    Eq.~(\ref{eqn:Regret2}).\\
     \ \ 4. Then the probability distribution of the actions for \\
     \ \ \ \ \ the next period, $p_i^{t}$ is defined as: \\
     \vspace{-.4cm}
     \begin{align}
     \begin{array}{l}
     p_i^{t}(\hat \alpha_i^{t-1})=\frac{1}{\mu}\mathbb{R}_i^{t-1}(\alpha_i^{t-1}, \hat \alpha_i^{t-1}),\\
     p_i^{t}(\alpha_i^{t-1})=1- p_i^{t}(\hat \alpha_i^{t-1}),
     \end{array}
     \label{eq:regret_finding_p}
     \end{align}
     \ \ \ \ \ where $\mu$ is a certain constant that is
     sufficiently large.\\
    \hline
  \end{tabular}
  }\vspace{-3mm}
\end{table}

\section{Simulation Results}\label{sec:simulation}\vspace{-1mm}

\begin{figure}[h]
\centering
\includegraphics[width=.4\textwidth]{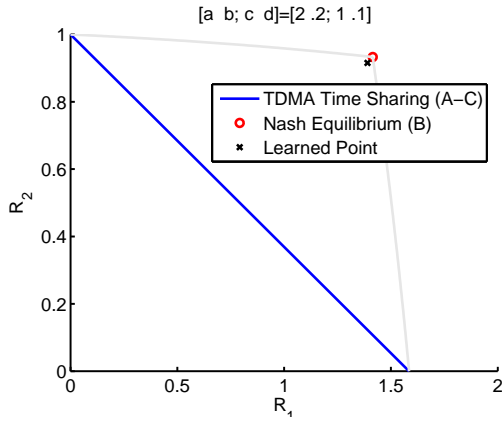}
\caption{Noise-limited: $2-$user case} \label{fig:2_noise}
\end{figure}

\begin{figure}[h]
\centering
\includegraphics[width=.4\textwidth]{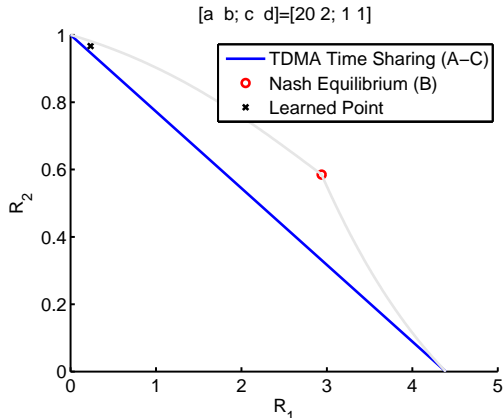}
\caption{Type II time-sharing (see Fig. \ref{fig:2Drateregion} (c))}
\label{fig:2_user2}
\end{figure}

\begin{figure}[h]
\centering
\includegraphics[width=.4\textwidth]{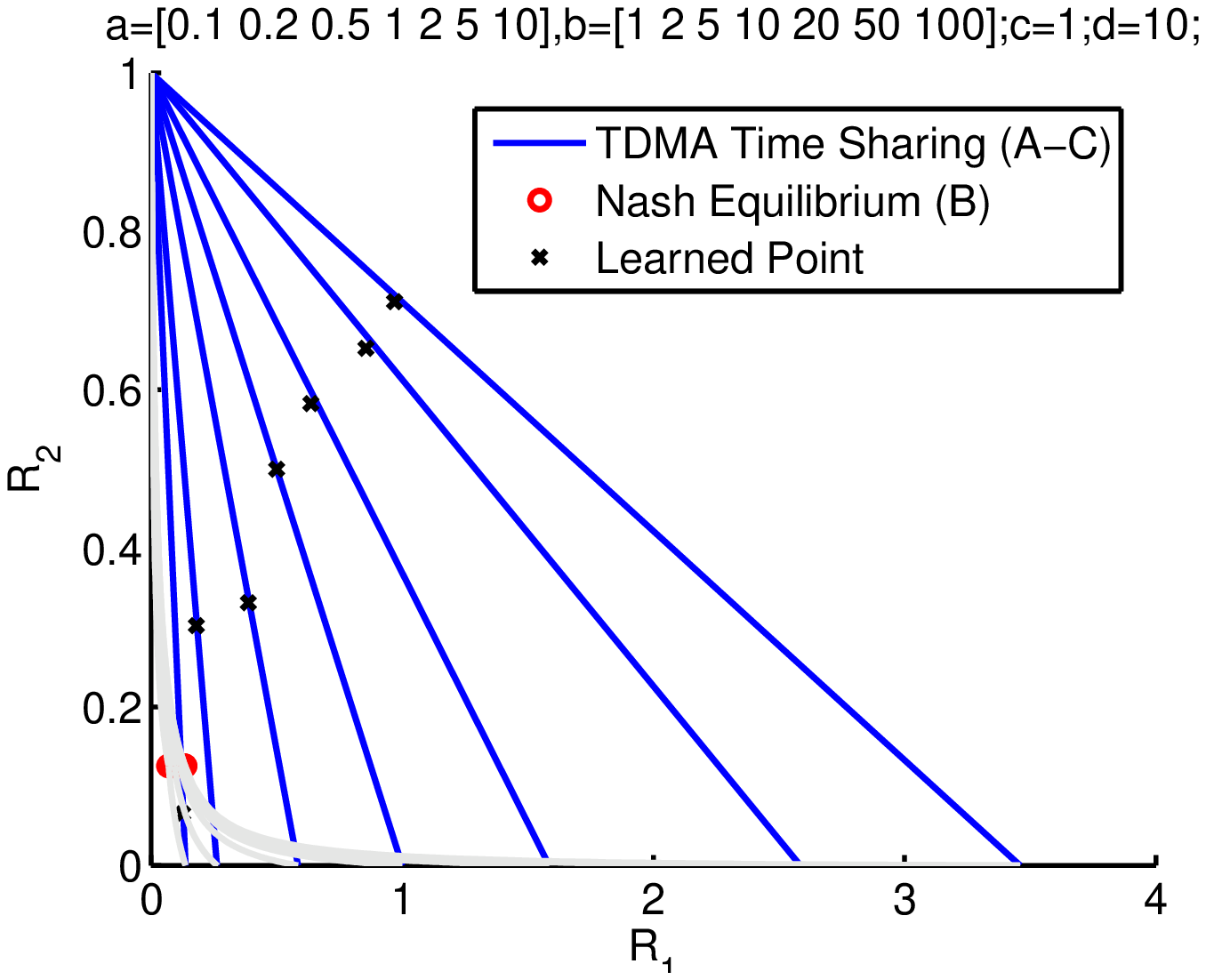}
\caption{Interference-limited: Type I time-sharing}
\label{fig:2_inter}
\end{figure}

To demonstrate the proposed scheme, we setup a $2-$user
interference channel where $P_{\max}=1$. In Fig.
\ref{fig:2_noise}, we show the crystallized rates region for the
noise-limited regime with $a=2,b=0.2,c=1$, and $d=0.1$. The learning
algorithm converges close to the Nash equilibrium, which means that
both users transmit with maximum power $P_{\max}$ all the
time. This corresponds to the case in Fig. \ref{fig:2Drateregion}
(a). In Fig. \ref{fig:2_user2}, we show the Type II time-sharing
case with $a=20, b=2, c=1$, and $d=1$. The algorithm converges to
$\theta_2=0.92$ and $\theta_3=0.08$, which means  the probability
that user 2 transmits alone is $0.92$, and the probability that both
users transmit with full power is $0.08$. This corresponds to the
case in Fig. \ref{fig:2Drateregion} (c). Finally, in Fig.
\ref{fig:2_inter}, we show the interference-limited regime with
$c=1, d=10$ as well as seven different instances of $a$ and $b$.
First, the Nash equilibriums exhibit much poorer performance than
the TDMA time-sharing lines. The proposed learning algorithm
converges to a point on the TDMA time-sharing lines, this
corresponds to the case in Fig. \ref{fig:2Drateregion} (d).
Moreover, the learning algorithm favors the weaker user.

\begin{figure}[h]
\centering
\includegraphics[width=.4\textwidth]{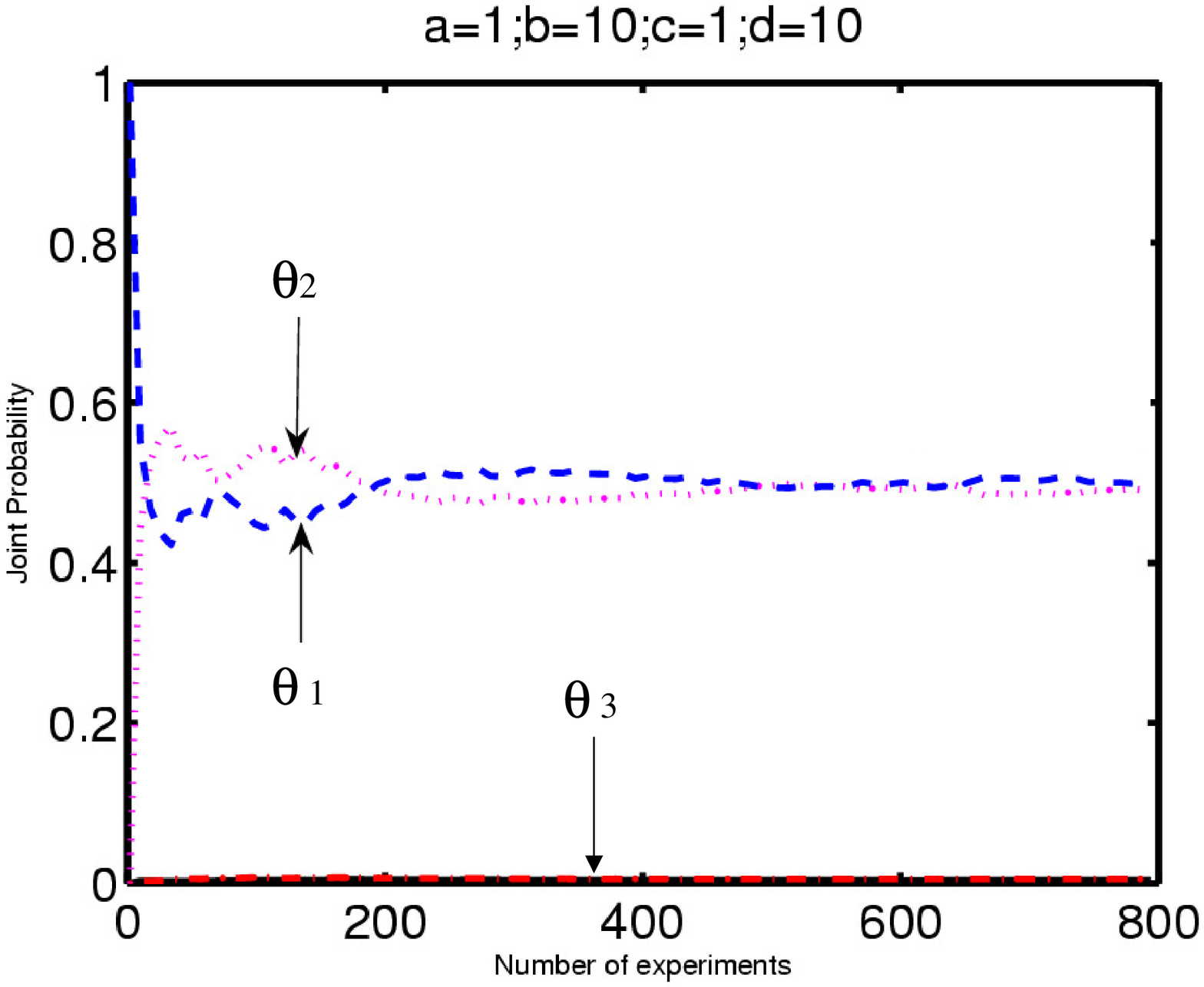}
\caption{Interference-limited: Type I time-sharing}
\label{fig:2_sharing}\vspace{-3mm}
\end{figure}

In Fig. \ref{fig:2_sharing}, we show the interference-limited case
with $a=1,b=10,c=1$, and $d=10$. Due to the symmetry, the learning
algorithm achieves probabilities of $0.5$, which means the two users
conduct equal time-sharing over the channel, where each transmits
solely at full power while the other is silent; and such two
transmission states happen equally 50\% of the time each. This
corresponds to the A-C time-sharing case in Fig. \ref{fig:2Drateregion} (d).

\section{Conclusion}\label{sec:conclusion}

Treating the interference as noise, the paper proposes a novel approach to
the rates region in the $n-$user interference channel, composed by the time-sharing
convex hull of $2^n-1$ corner points achieved through On/Off binary power control.
The resulting rates region is denoted crystallized rates region. It then applies
the concept of correlated equilibrium from game theory to form the convex hull of
the crystallized region. An example in applying these concepts for the $2-$user case,
the paper considered a mechanism design approach
to design the Vickrey-Clarke-Groves auction utility function. The regret-matching
algorithm is used to converge to the solution point on the correlated equilibrium
set, to which subsequently simulation was presented.

%{\bf Zhu: please flip the $R_1$ and $R_2$ on the x-y axis in the simulation figures, thanks}

%In this paper, we propose the concept of time sharing binary power
%control to form a {\em crystallized} rates region, which renders
%utility maximization more tractable. Then, we use correlated
%equilibrium to approach the problem due to its convexity nature of
%the solution set. To link the crystallized rates region with the
%correlated equilibrium set, we employ the mechanism design. Finally,
%non-regret learning is used to achieve the correlated equilibrium
%set (i.e. the boundary of crystallized rates region) in a
%distributive manner. We show through simulation the gradual tendency
%towards time-sharing as interference gain increases in 2-user and
%3-user scenarios. \vspace{-3mm}

%\begin{figure}[h]
%\centering
%\includegraphics[width=.33\textwidth]{3user_interference_learning.eps}
%\caption{Interference Dominated: 3 User Case}
%\label{fig:3_inter}\vspace{-5mm}
%\end{figure}

% ------------------------------ BIBLIOGRAPHY ----------------------------------
\bibliographystyle{IEEE}

\end{document}